\documentstyle[sprocl,epsfig]{article}

\def\be{\begin{equation}}
\def\ee{\end{equation}}
\def\bea{\begin{eqnarray}}
\def\eea{\end{eqnarray}}
\newcommand{\lsim}{\raisebox{-0.13cm}{~\shortstack{$<$ \\[-0.07cm] $\sim$}}~}

\begin{document}
\vspace{-2truecm}
\begin{flushright}
PM/99--36
\end{flushright}

\title{A DIRECT RECONSTRUCTION OF \\ 
THE GAUGINO PARAMETERS WITH PHASES} 

\author{ J.-L. KNEUR~\footnote{Speaker}  and G. MOULTAKA }

\address{Physique Math\'ematique et Th\'eorique, UMR No 5825--CNRS, \\
Universit\'e Montpellier II, F--34095 Montpellier Cedex 5, France}


\maketitle\abstracts{
A simple algebraic algorithm is described to recover 
the (complex valued) gaugino/Higgsino basic parameters 
$\mu \equiv \vert \mu\vert e^{i \phi_\mu}$, 
$M_1 \equiv \vert M_1 \vert  e^{i \phi_{M_1}}$,  
$M_2\equiv |M_2|$, 
directly in terms of (a minimal set of) neutralino and chargino 
masses, accurately measurable
at future linear collider energies.  
An ambiguity in the $M_1$ reconstruction   
can be resolved 
by measuring the corresponding $e^+e^- \to  \chi^0_1 \chi^0_2$
cross-section. This approach should simplify the
determination of allowed ranges of the gaugino parameters, 
in particular if only a partial set of gaugino masses were
measured.}   
\section{Introduction and Motivations}
The unconstrained
Minimal Supersymmetric extension 
of the Standard Model \linebreak (MSSM)~\cite{R1} 
involves a large
number of arbitrary parameters, which furthermore may be
in general complex valued, adding  new sources of
CP violation. 
Relatively large phases in the ``flavour-blind"
gaugino and/or
Higgs sector
are not excluded by present constraints~\cite{edmexp,edmth}, 
and may lead to drastic changes in the
phenomenology of the Higgses and 
gauginos~\cite{choi}$^-$\cite{Bargeretal}, 
affecting the reconstruction from 
data of the structure of the
SUSY and soft-SUSY breaking Lagrangian.  
We shall report here on a specific construction 
and algorithm~\cite{inoinv1,inoinv2} to obtain the
basic gaugino parameters 
in direct analytic form in terms of physical
masses, including possible non-zero phases~\cite{inoinv2}.
We illustrate in particular how this
``dediagonalisation" approach can reveal in an 
easy way the non-trivial correlations 
among the chargino and neutralino sectors,   
exhibiting directly e.g. allowed domains for the neutralino masses, once
some of the parameters of the chargino sector are determined~\cite{choi}, 
or vice-versa.
\section{Extracting $\mu$, $M_2$ and $M_1$ from physical masses}
 Without loss of
generality, in the unconstrained MSSM (i.e. no
universality of the gaugino mass terms is assumed), we  
choose a phase convention such that the soft  susy breaking
$SU(2)_L$ Wino mass $M_2$, is real.
The chargino
mass matrix thus reads  
\be
\label{Mchargino}
M_C = \left(
  \begin{array}{cc} M_2  & \sqrt 2 m_W \sin\beta  \\
              \sqrt 2 m_W \cos\beta  & \vert \mu \vert e^{i \Phi_\mu}  
\end{array} 
\right)
\ee 
and the (squared) chargino masses  
are obtained as the eigenvalues of the
$M^\dagger_C M_C $ matrix.
The neutralino mass matrix with the relevant phases reads
\be
\label{Mneutralino}
M = \left(
  \begin{array}{cccc} |M_1| e^{i\Phi_{M_1}} 
& 0 & -m_Z s_W \cos\beta & m_Z s_W \sin\beta  \\
  0 & M_2 &  m_Z c_W \cos\beta & -m_Z c_W \sin\beta  \\
 -m_Z s_W \cos\beta & m_Z c_W \cos\beta & 0 & -|\mu| e^{i\Phi_\mu} \\
m_Z s_W \sin\beta & -m_Z c_W \sin\beta & -|\mu| e^{i\Phi_\mu} & 0 
\end{array} \right)
\ee 
where $|M_1| e^{i\Phi_{M_1}}$  is the soft supersymmetry
breaking $U(1)_Y$ Bino mass,
while $|\mu| e^{i\Phi_\mu}$ parametrizes
the supersymmetry conserving mixing 
of the two Higgs doublets. 
\subsection{Chargino de-diagonalization}
Assuming as input $\tan\beta$, 
the two physical chargino masses and one mixing angle (say $\phi_L$ for
definiteness), one can derive straightforwardly   
``inverted" expressions for $\vert\mu\vert$, $\Phi_\mu$ and $M_2$
directly in terms of physical parameters~\cite{inoinv2},~\footnote{A method 
to extract the same parameters from
chargino pair production is described in ref.~\cite{choi}.}:
\be
\label{mum2}
|\mu| (M_2) = \left[\frac{1}{2}(\Sigma -2 m^2_W (1 +(-) \cos 2\beta
-(+) \Delta \cos(2\phi_L) )\right]^{1/2}
\ee
%($\Sigma\:, \Delta \equiv 
%M^2_{\chi_2} \pm M^2_{\chi_1}$) and
\be
\label{cphmu}
\cos(\Phi_\mu) = 1 -{\frac{M^2_{\chi_1} M^2_{\chi_2}
-( M_2 \: |\mu| - m^2_W\sin 2\beta)^2}
{2 m^2_W \: M_2 \: |\mu| \;\sin 2 \beta } }\; .
\ee
where $\Sigma\:, \Delta \equiv 
M^2_{\chi_2} \pm M^2_{\chi_1}$. Now, clearly not any input 
$M_{\chi_1}$, $M_{\chi_2}$ will lead to 
$|\mu|$, $\cos\Phi_\mu$, $M_2$ values consistent 
with the obvious constraints $M_2, |\mu| \geq 0$, 
and $| \cos\Phi_\mu | \leq 1$. Depending on the actual values
of $M_{\chi_1}$, $M_{\chi_2}$, 
this can give relatively strong 
constraints, as illustrated below.  
\subsection{Neutralino sector}
We now determine $\vert M_1 \vert$ and $\Phi_{M_1}$ 
for given  $M_2$, $\vert\mu\vert$, 
$\Phi_\mu$, $\tan  \beta$
and two (arbitrary) input neutralino masses $M_{N_1}$, $M_{N_2}$\footnote{ 
For definiteness $M_{N_1}$, $M_{N_2}$
are assumed to be the lightest and next to lightest.
It should be clear, however,  that any two masses among the four
can be equivalently used as input.}.
Relatively simple  
explicit expressions for $\vert M_1 \vert$ and $\Phi_{M_1}$  
are obtained~\cite{inoinv1,inoinv2} 
by considering the four invariants of the hermitian matrix
$M^\dagger M$,
given by the different coefficients  
of the characteristic polynomial 
$ det(M^\dagger M -M^2_N I) $ in $M^2_N$.
However, the resulting system is quadratic in $M_1$, which leads to an
intrinsic twofold ambiguity in its determination~\footnote{An alternative
method to extract $\mu$, $M_1$, $M_2$ 
uniquely at linear collider energies was
recently investigated in ref.~\cite{Bargeretal}, 
however using the full set of neutralino masses as data.}.
Also, quite similarly to the simpler
chargino case discussed above, non-trivial constraints among the 
neutralino masses arise, simply 
because the system
cannot always have a solution consistent with
$|\cos\Phi_{M_1}|,
|\sin\Phi_{M_1}| \leq 1$, for {\em any} input 
$M_{N_1}$, $M_{N_2}$. 
\subsection{Illustration of reconstruction from a complete set of data}
In fig. 1a,b are plotted the output moduli and  
phases respectively, as functions of the  
chargino mixing angle $\phi_L$, for typical input chargino and neutralino 
masses.
The announced consistency correlations among the chargino and
neutralino masses appear explicitly here in the fact that 
the $\phi_L$ domain where 
both sectors can consistently exist is quite narrow 
($0.37 \lsim \phi_L \lsim 0.51$; $0.9 \lsim \phi_L \lsim 1.$ (rad.)). 
Moreover, relatively moderate
changes in the input mass values may easily result in narrower
or even empty solution zones: changing e.g. 
$M_{\chi_1}$ only from 80 to 100 GeV, for the same values of the other
input parameters, gives no consistent $M_1$, $M_2$, $\mu$.  
%%%%%%%%%%%%%%%%%%%%%%%%%%%%%%%%%%%%%%%%%%%%%%%%%%%%%%%%%%%%%%%%%%%%%%%
% FIG. 1
\begin{figure}[t]
%\vspace{-4.cm}
\epsfxsize=5.in
\epsfysize=7.in
\epsffile[160 0 860 700]{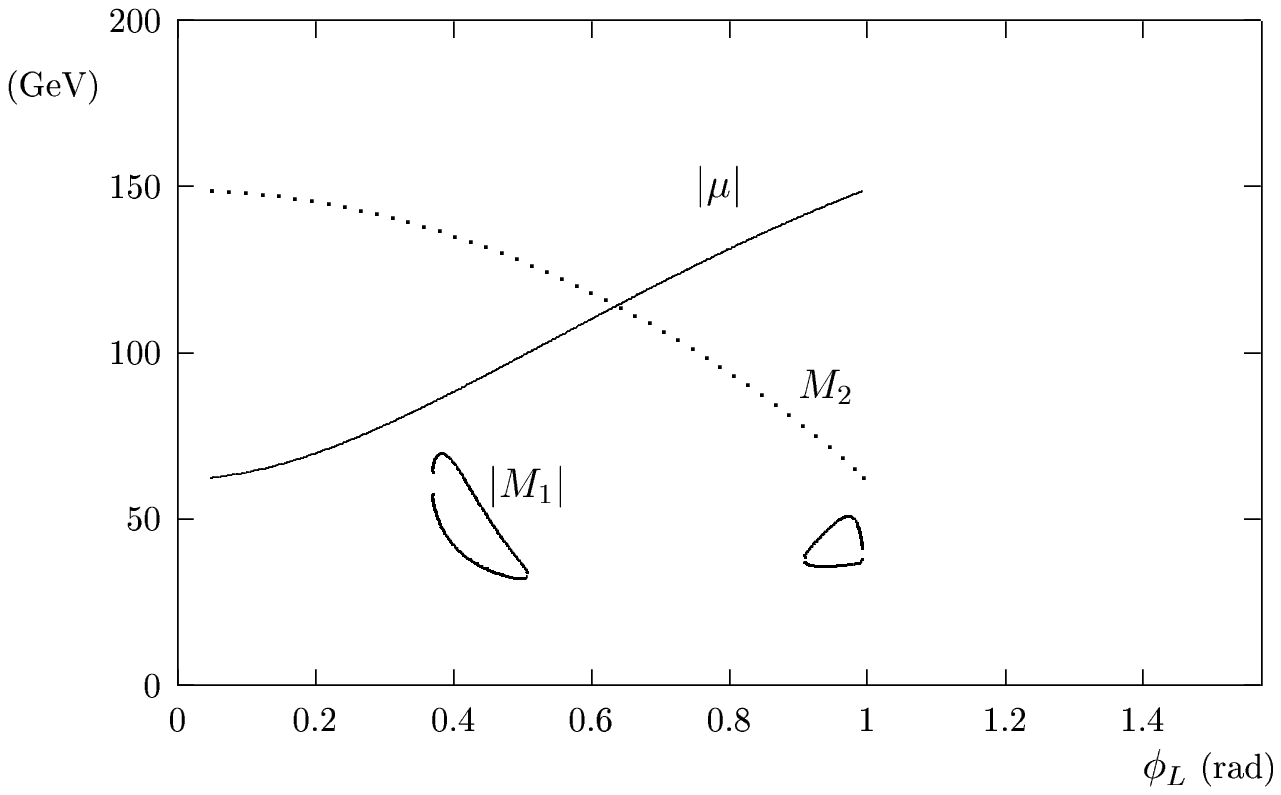}
\vspace{-18.2cm}
\epsfxsize=5.in
\epsfysize=7.in
\epsffile[-225 0 475 700]{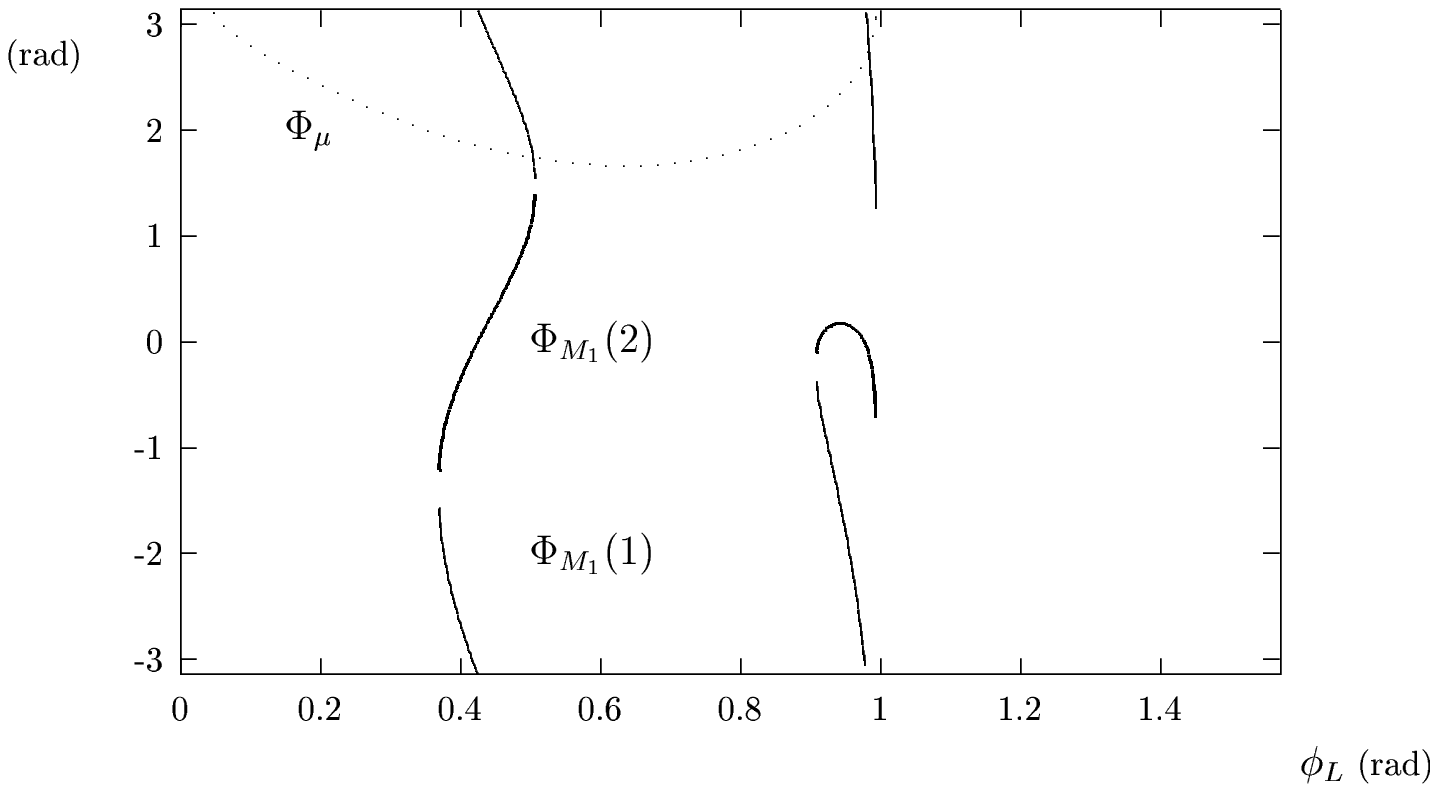}
\vspace{-12.cm}
\epsfxsize=15.cm
\epsffile[50 50 700 700]{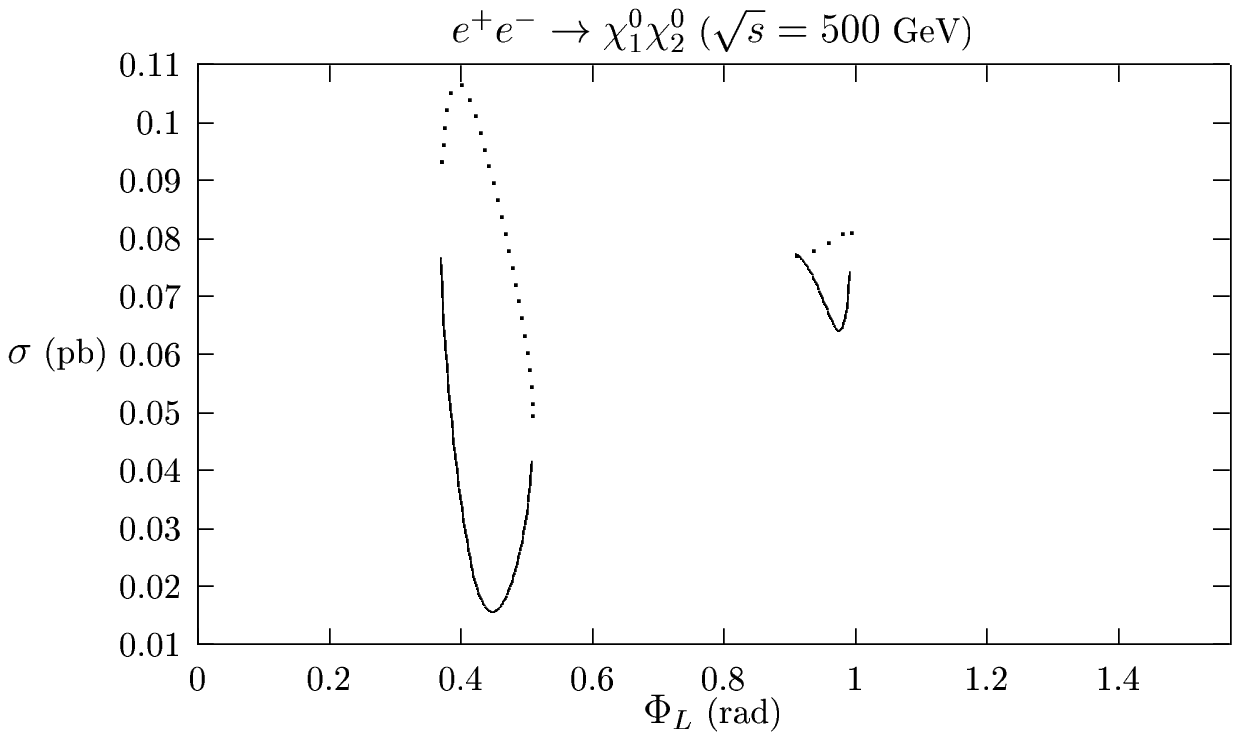}
\vspace{-10.5cm}
% autre position:
%\begin{figure}
%\epsfxsize=15.cm
%\epsffile[100 100 700 700]{fig3_mod.ps}
%\vspace{-9.5cm}
%\epsfxsize=15.cm
%\epsffile[100 100 700 700]{fig3_phi.ps}
%\vspace{-9.5cm}
%\epsfxsize=15.cm
%\epsffile[100 100 700 700]{neutralino-500-320-300.ps}
%\vspace{-10.5cm}
\caption{ (a) (upper left) reconstruction of $|M_1|$, $M_2$, $|\mu|$ ; 
(b) (upper right) reconstruction of $\Phi_\mu$, $\Phi_{M_1}$. 
Physical input sample: $M_{\chi^+_1} =$ 80 GeV, $M_{\chi^+_2} =$ 180
GeV,  $M_{N_1}=40$, $M_{N_2}=80$;  $\tan\beta=2$; (c) (bottom fig.)
corresponding $\chi^0_1, \chi^0_2$ production
cross-section for $\sqrt{s}= 500$ GeV, $m_{\tilde e_L, \tilde e_R}=$
320, 300 GeV .} 
\end{figure}
%%%%%%%%%%%%%%%%%%%%%%%%%%%%%%%%%%%%%%%%%%%%%%%%%%%%%%%%%%%%%%%%%%%%%%
%
\subsection{The $e^+e^- \to \chi^0_i\chi^0_j$ cross-section with
phases} 
Assuming non-zero $M_1$, $\mu$ phases, 
it is crucial to analyze the changes implied
in the relevant gaugino production 
cross-sections\cite{bartl1}$^-$\cite{lcrep,choi,inoinv2}.   
Here we illustrate
the neutralino 1,2 pair production 
at future linear $e^+e^-$ collider,
plotted in fig. 1c 
as a function of the
chargino, neutralino masses  and chargino mixing angle,  
for the same choice of parameters corresponding to fig. 1a,b and 
one choice of the selectron masses. 
One first observes the generically rather important
sensitivity of this cross-section to the phases~\cite{Petcov} 
(the plots in fig. 1c
are functions of the mixing angle $\phi_L$, which corresponds
for fixed chargino and neutralino masses to varying $\Phi_\mu$ 
and $\Phi_{M_1}$, see 
corresponding fig. 1b). 
Moreover, 
and quite generically, the 
difference in the
cross-section values for the two different $M_1$ solutions of fig.1
should easily resolve this two-fold ambiguity,  
provided the cross-section will be large enough to be measured with a
reasonable accuracy.
Furthermore,  as a by-product of
the inversion procedure, one can obtain direct correlations among
e.g. the chargino masses and pair production, and the neutralino pair
production cross-sections.
\section{Constraints on basic parameter space from incomplete data}
Let us consider
now a less optimistic situation where only a partial knowledge of the physical
input masses is assumed, and illustrate the 
kind of information that can be
retrieved in this case. Accordingly, in fig. 2 we suppose that none of
the neutralino masses are known, while 
the two chargino masses and $\tan \beta$ are fixed,
and $\phi_L$ varies between  
$\phi_L \simeq 0.37$ and $\phi_L \simeq 0.51$ (rad), corresponding 
to the first consistent zone in fig. 1a,b. 
Consistency directly implies 
the pattern of correlation among the only possible 
physical neutralino masses,
represented by the dotted regions (``butterflies") in the figure. 
More precisely, a given set of the four neutralino masses is consistent
only if any pair $(M_{N_i}, M_{N_j})$ of these masses corresponds to a
point lying on one of the ``butterflies". 
From this requirement one can actually 
draw definite consequences such as the fact that each of the three
allowed branches (along say the y-axis) can host only {\sl one}
pair $(M_{N_i}, M_{N_j})$, etc.
When $\tan\beta$ increases, the
dotted  butterflies are simply moving up or down along the diagonal 
$M_{N_i}=M_{N_j}$ line. 
We emphasize that within this algorithm,
such correlations are very simply obtained
from scanning over arbitrary values of the two input neutralino masses,
checking the consistency relations only once for each $M_{N_i}$, $M_{N_j}$,
(and eventually $\phi_L$) input. 
Another example of possible application, 
that we do not illustrate here,
is the direct determination of the 
allowed domains~\cite{inoinv2} of the $\mu$ or $M_1$ phases, 
when a given range of
the chargino and/or neutralino masses will be delineated by data.   
%%%%%%%%%%%%%%%%%%%%%%%%%%%%%%%%%%%%%%%%%%%%%%%%%%%%%%%%%%%%%%%%%%%%%
% FIG 2
\begin{figure}
%\vspace{-2cm}
\epsfxsize=12cm
\epsffile[0 0 700 700]{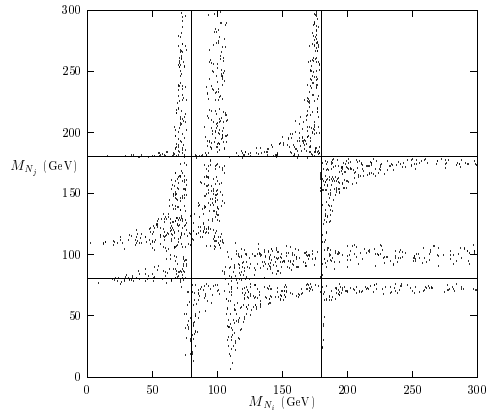}
\vspace{-6.5cm}
\caption{ \label{fig5-1} Correlations between arbitrary
neutralino masses from consistency of the inversion. Input 
choice: $M_{\chi^+_1} =$ 80 GeV, $M_{\chi^+_2} =$ 180
GeV,  0.37 $< \phi_L(rad) <$ 0.52;  $\tan\beta= $2.}
\end{figure}
%%%%%%%%%%%%%%%%%%%%%%%%%%%%%%%%%%%%%%%%%
%
\section{Conclusions}
A simple algebraic algorithm is derived to reconstruct 
the (unconstrained) gaugino sector
Lagrangian parameters $\mu \equiv |\mu| e^{i\Phi_\mu}$, 
$M_2\equiv |M_2|$ and 
$M_1 \equiv |M_1| e^{i\Phi_{M_1}}$, directly from the physical
chargino and (some of the) neutralino masses.  
Our approach 
should be useful in particular in 
the case where only a subset of the minimal required input is
available, where it exhibits in a more direct way the
non trivial correlations among the physical chargino and neutralino
parameters. These correlations exist even when the
maximal possible phase freedom of the unconstrained MSSM parameter space
is considered, and  may be hidden or very tedious to extract
in the more standard approach of systematic scanning 
over the basic parameters and fitting the data. 
Note finally that the inclusion of
realistic mass measurement errors, given 
the accuracy level expected at a future 
linear collider~\cite{lcrep,Bargeretal},
affects only very scarcely e.g. the shape of 
figs. 1, 2 and our reconstruction in general.  

\end{document}